\documentclass[%
 reprint,
superscriptaddress,
nofootinbib,
pra,
floatfix,
]{revtex4-1}

\usepackage[english]{babel}
\usepackage[utf8]{inputenc} 
\usepackage{graphicx}
\usepackage[hidelinks]{hyperref}
\usepackage{physics}
\usepackage{bropd}
\usepackage{siunitx}

\usepackage{microtype}

\begin{document}

\preprint{000/000-000}

% Title
\title{Efficient polarization of high-angular-momentum systems}

\author{Simon M. Rochester}
\affiliation{Rochester Scientific, LLC, El Cerrito, CA 94530, USA}

\author{Konrad Szymański}
\affiliation{Instytut Fizyki, Uniwersytet Jagielloński, ul. Łojasiewicza 11, 30--348 Kraków, Poland}

\author{Mark Raizen}
\affiliation{Department of Physics, University of Texas at Austin, Austin, TX 78712, USA }

\author{Szymon Pustelny}
\affiliation{Instytut Fizyki, Uniwersytet Jagielloński, ul. Łojasiewicza 11, 30--348 Kraków, Poland}

\author{Marcis Auzinsh}%M\={a}rcis Auzi\c{n}š}
\affiliation{Laser Centre, University of Latvia, Rainis Boulevard 19, LV--1586 Riga, Latvia}%Faculty of Physics and Mathematics, Latvijas Universit\={a}te, Ze\c{l}\c{l}u iela 8, R\={\i}ga, LV--1002}

\author{Dmitry Budker}
\affiliation{Helmholtz Institut, Johannes Gutenberg Universität, 55099 Mainz, Germany}
\affiliation{Department of Physics, University of California, Berkeley, CA 94720-7300, USA}
\affiliation{Nuclear Science Division, Lawrence Berkeley National Laboratory, Berkeley, CA 94720, USA}

\date{\today}% 
\begin{abstract}
We propose methods of optical pumping that are applicable to open, high-angular-momentum transitions in atoms and molecules, for which conventional optical pumping would lead to significant population loss. Instead of applying circularly polarized cw light, as in conventional optical pumping, we propose to use techniques for coherent population transfer (e.g., adiabatic fast passage) to arrange the atoms so as to increase the entropy removed from the system with each spontaneous decay from the upper state. This minimizes the number of spontaneous-emission events required to produce a stretched state, thus reducing the population loss due to decay to other states. To produce a stretched state in a manifold with angular momentum $J$, conventional optical pumping requires about $2J$ spontaneous decays per atom; one of our proposed methods reduces this to about $\log_2 2J$, while another of the methods reduces it to about one spontaneous decay, independent of $J$.
\end{abstract}
\maketitle

\section{Introduction}
\label{sec:intro}

Although it has been employed for over 50 years~\cite{Kastler1950}, optical polarization of atoms and molecules is steadily gaining new applications in science and technology~\cite{Happer1972,Auzinsh1995,Auzinsh2010,Happer2010}. For some of these applications, it may be valuable to revisit established techniques. 

One such technique is the generation of a so-called stretched (maximal projection) state, i.e., a state in which only the Zeeman sublevel with the highest or lowest $m$ quantum number (the ``end state'') is occupied. Due to its simplicity, the stretched state is often used as a starting point for quantum optics, spin squeezing, and sensing experiments. In certain experimental situations, the stretched state has other advantages, for instance, being immune to relaxation due to spin-exchange collisions (see Ref.~\cite{Scholtes2011} and references therein). Stretched states can also be used to cool atoms or molecules by controlling their motion through the interaction of their magnetic moments with a spatially varying magnetic field (magneto-optical cooling)~\cite{Raizen2014}. It is also possible to use such states in laser control of chemical dynamics, as the efficiency of some processes depends on the relative polarization of the reacting compounds~\cite{Simons1987}.

A particular application of the stretched state is the generation of spin squeezing (a redistribution of uncertainties from one spin component to another) induced by a mechanism known as orientation-to-alignment-conversion (OAC)~\cite{Rochester2012_OAC}. In this process, atoms in a stretched state interact with an electric field and the state evolves into one in which the spin-projection uncertainty in a certain direction is diminished at the expense of uncertainty in another direction. Spin squeezing has been investigated as a means to enhance the sensitivity of precision measurements, such as magnetometry (see Ref.~\cite{Sewell2012} and references therein).

For OAC squeezing, the best results are predicted in systems having high total angular momentum~\cite{Rochester2012_OAC}, such as molecules, where rotational excited states with total angular momentum as high as $J=100$ are routinely observed. This raises the question of how to efficiently optically polarize high-angular-momentum systems. There are papers describing experiments on high-angular-momentum molecules~\cite{Auzinsh2012,Auzinsh2008}. No efficient method of direct optical pumping designed for such systems has been found yet, although indirect methods like spin-exchange optical pumping are in some cases applicable. 

The creation of a stretched state by conventional optical pumping involves using circularly polarized light to transfer atoms\footnote{Or molecules---we will use atoms as a generic term.} from an initially populated lower state to an unpopulated upper state, from which they subsequently decay. Multiple cycles of absorption followed by spontaneous decay tend to result in a stretched state (at least for a $J\to J'=J$ transition, the simplest case). This can generate significant polarization, but possibly at the expense of population: unless a closed transition is employed, atoms may decay to states other than the lower state of interest and be lost. This is especially of concern for transitions between high-$J$ states, both because the branching ratios for spontaneous emission to the desired states are often low, and because many (on the order of $2J$) spontaneous emission cycles are required for polarization. Considering for example a $J\to J'=J$ transition, in the worst case all of the lower-state sublevels except for the end state are depleted without appreciably increasing the population of the end state, resulting in a stretched state with population a fraction $\approx\!1/(2J)$ of the total initial lower-state population. Population loss may be mitigated by using additional repump light beams, or by broadening the optical fields or transitions so that one field may serve as both pump and repump. On the other hand, this may be difficult or ineffective if there are many possible decay channels from the upper state.

Minimizing spontaneous emission can also be important even if there is no loss of population.  For example, the method of magneto-optical cooling~\cite{Raizen2014} requires cycles of optical pumping together with magnetic kicks. The ultimate temperature of the atoms is limited by the number of spontaneous emission events, so there is clear benefit in minimizing their number. 

There are well-known methods for transferring populations between sublevels coherently, i.e., via absorption and stimulated emission only, without depending on spontaneous emission. For example, lower- and upper-state populations may be swapped through the technique of adiabatic fast passage (AFP), in which an optical pulse is applied with its frequency swept through resonance. The frequency sweep must satisfy the adiabatic condition that the timescale of the sweep must be much shorter than the spontaneous-decay lifetime, but much longer than the Rabi-oscillation period. On the other hand, the pulse parameters do not need to be tuned for a specific transition strength, as would be required for a standard $\pi$ pulse. Thus, a single AFP pulse can be used to simultaneously swap the populations on each $m\to m'$ sublevel transition of a $J\to J'$ transition manifold, even though the transition strengths for each individual $m\to m'$ transition are different. The adiabaticity requirement means that this method is limited to cases in which the lifetimes are sufficiently long and light intensities can be made sufficiently high.

Another approach is stimulated Raman adiabatic passage (STIRAP), which uses a two-pulse sequence to transfer atoms from a populated state to an initially unpopulated state with the aid of an intermediate state, without ever developing an appreciable population in the intermediate state~\cite{Bergmann1998}. Thus spontaneous decay can be eliminated, even if the lifetime of the intermediate state is short.

The drawback of coherent-population-transfer methods in the context of our current discussion is that, by themselves, they cannot be used to increase atomic polarization---in the sense that they cannot cause atoms that are initially in different sublevels to be combined into the same sublevel. This follows from the fact that coherent processes involve unitary (reversible) evolution, as opposed to irreversible evolution due to a relaxation process like spontaneous decay. (Here we are considering a semiclassical model in which the atoms are treated quantum mechanically, the applied light fields are treated classically, and interactions with other systems, such as the vacuum field modes, are treated phenomenologically as relaxation processes.) 

The effect of reversible evolution can be most easily seen for the case of an initial state with no coherences between the sublevels that undergoes unitary evolution to a final state that also has no coherences. A state with no coherences is described by a diagonal density matrix, with the sublevel populations along the diagonal. Unitary evolution is represented by a unitary transformation of the density matrix. A unitary transformation is known to preserve the eigenvalues of a matrix, which for a diagonal matrix are equivalent to the diagonal entries. Thus, in this case, the sublevel populations cannot be altered by the coherent evolution, although their order can be changed (i.e., populations can be swapped between sublevels).

In general, a unitary process may take an initial state with no coherences to one that has coherences. In this case one can use the general result that the diagonal entries of any Hermitian matrix are bounded by its smallest and largest eigenvalues. These eigenvalues are given by the smallest and largest populations of the initial (diagonal) density matrix, and are not changed by the unitary transformation. Thus, even though the populations may change, there can never be any population smaller than the initial smallest sublevel population or larger than the initial largest population. In particular, if the initial state consists of a lower state with equally populated sublevels and an unpopulated upper state, no lower-state sublevel population can increase beyond its initial value.

Another way to describe this principle is in terms of the entropy of the atomic system. Because, in the semiclassical approximation, coherent processes describe the evolution of a closed quantum system, they cannot change the entropy. To see this, note that the von Neumann entropy per atom (which we express in units of the Boltzmann constant) for a system with density matrix $\rho$ is defined as the expectation value~\cite{Happer2010,Haroche2006}
\begin{equation}
	S_\text{atom}=-\ev{\ln\rho}=-\tr(\rho\ln\rho)=-\sum_i\lambda_i\ln\lambda_i,
\end{equation}
where we have taken the trace in the diagonal basis of $\rho$, so that the matrix logarithm reduces to an expression in terms of the eigenvalues $\lambda_i$ of $\rho$ \cite{Tarantola2006Elements}. We see that the entropy of a pure state---which has a single nonzero eigenvalue $\lambda=1$---is zero, while the entropy of a mixed state is positive. Thus, increasing the polarization corresponds to reducing the entropy. But since a unitary transformation cannot change the eigenvalues, it cannot reduce the entropy of the system.

A relaxation process, on the other hand, can reduce the entropy of the atomic system. This is because relaxation processes describe the interaction between the atomic system and an external system, and entropy can be transferred from one to the other. In the case of spontaneous emission, entropy is carried away into the electromagnetic-field modes. (In a fully quantum-mechanical treatment, the laser field is also a dynamical system that can take up entropy as a result of relaxation or decoherence processes, even in the absence of spontaneous decay \cite{Metcalf2008}; this effect is negligible in our consideration of optical pumping.)

% \SR{If the laser field is treated quantum mechanically, it too can take up entropy, and Metcalf claims that this can be used to perform momentum cooling without SE~\cite{Metcalf2008,Corder2015}. That doesn't seem relevant here, though, since we don't have to use a full quantum description to accurately describe OP. I would rather not cite Metcalf unless we can say why the laser field entropy is relevant in his case but not in ours.}

While coherent methods alone cannot be used to create a stretched state (other than by simply depleting all sublevels except for the end state, which is no better than the worst-case efficiency of conventional optical pumping), they can be used to increase the efficiency of optical pumping. Conventional optical pumping uses, in general, far more spontaneous-decay cycles per atom (on the order of the number of ground-state sublevels) than are necessary to transfer the initial entropy out of the atomic system. To see how many cycles are actually required, consider the entropy of an initially unpolarized state with angular momentum $J$. There are $2J+1$ sublevels, each with a population of $1/(2J+1)$, and the entropy is
\begin{equation}
	S_\text{atom}^\text{\,initial}=-\sum_{m=-J}^{J}\frac{1}{2J+1}\ln\frac{1}{2J+1}=\ln(2J+1).
\end{equation}
Thus the entropy to be removed scales logarithmically with the number of sublevels.

The production of entropy in spontaneous emission was discussed in Ref.~\cite{vanEnk1992}. The entropy increase of the optical field due to a single spontaneously emitted photon was found to be $1-\ln f$, where $f$ is the total number of spontaneously emitted photons per mode. The number of modes can be estimated as $\ell^2/\lambda_0^2$, where $\ell$ is the linear dimension of the sample volume and $\lambda_0$ is the transition wavelength. (As discussed in Ref.~\cite{vanEnk1992}, this is because the photons emitted in a given time interval are contained in an expanding spherical shell, and the entropy of these photons is constant after the photons have left the medium volume.) Suppose that an atomic population $p$ decays in a short time, resulting in $p$ spontaneously emitted photons. Then $f\approx p/(\ell^2/\lambda_0^2)$. Multiplying the expression for the entropy for one photon by $p$, we have, as an order-of-magnitude estimate for the total field-entropy increase,
\begin{equation}\label{eq:optS}
	S_\text{optical} = p\br{1-\ln f} \approx p\br{1-\ln p + \ln\frac{\ell^2}{\lambda_0^2}}.
\end{equation}
Typically $\ell\gg\lambda_0$, so the geometrical term is large, and generally dominates. However, for large $J$ (and $J'$), the population $p$ that decays from any one sublevel becomes small. In this case the leading term in Eq.~\eqref{eq:optS} is $-p\ln p$. For conventional optical pumping, there are ${\sim}2J$ sublevels, each with initial population $p\approx1/(2J)$; these populations each undergo spontaneous decay about $J$ times. Thus the leading term in the photon entropy is approximately $J\ln2J$, larger than the simultaneous reduction in atom entropy, ${\sim}\ln2J$. On the other hand, if there were an optical-pumping method in which each sublevel population underwent just one spontaneous decay while being transferred to the end state, the leading term would scale as $\sim\!\ln2J$, matching the scaling of the atom-entropy reduction. Thus, one decay per atom is the minimum number of spontaneous decays allowed by the second law of thermodynamics when creating a pure state (up to the accuracy of our order-of-magnitude estimates).

To improve the efficiency of optical pumping, we can arrange the atoms among the sublevels, using a coherent population-transfer method, to increase the entropy removed from the system by each spontaneous decay. Here we propose two schemes to achieve this. In the first, the sublevel populations are ``folded'' in half prior to each spontaneous decay. This can be achieved by transferring atoms between the sublevels of a ground $J$ manifold and an upper $J'$ manifold using AFP. Or, when the upper-state lifetime is short, it may be advantageous to implement the method using an additional shelving state. This folding scheme reduces the average number of spontaneous decays per atom from ${\sim}J$, obtained in conventional optical pumping, to ${\sim}\log_22J$. The second scheme, which uses two shelving states, can further reduce the average number of spontaneous decays per atom to one, which as discussed above is the minimum possible number for producing a completely polarized state. However, there is a trade-off when using this scheme: even though each atom is required to decay just once, only a fraction $1/(2J+1)$ of the population decays in each step, so that ${\sim}2J$ sequential decays are needed to complete the sequence. Thus this scheme may not be feasible if the upper-state lifetime is long and the total interaction time is limited.

\section{Description of proposed methods}
\label{sec:description}

\subsection{Folding method for $J\to J'=J$ systems}

\begin{figure}
\includegraphics{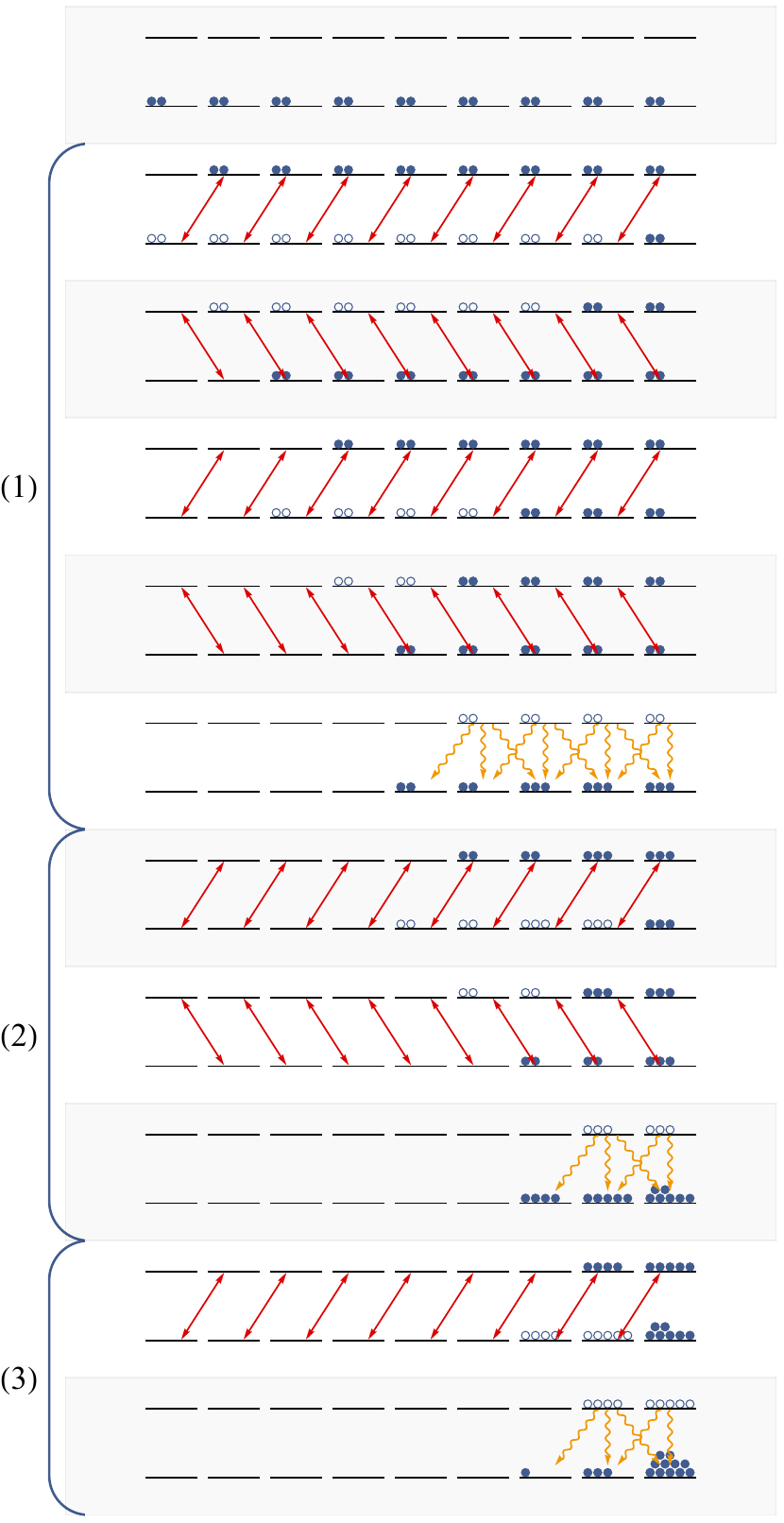}
\caption{\label{PulseLevelDiagrams44}Pulse sequence and resulting sublevel populations, obtained from the numerical model described in Sec.~\ref{sec:NumericalModel}, for the folding method applied to a $J=4\to J'=4$ system. The populations before each pulse are shown as open circles, and after each pulse as closed circles. The top diagram shows the initial state (unpolarized ground state and unpopulated upper state). Three iterations of the population folding procedure are applied, each followed by spontaneous decay of the upper state. In iteration (1), four ($=J$) AFP pulses are applied, with polarizations $\sigma^+$, $\sigma^-$, $\sigma^+$, $\sigma^-$. In iteration (2), two ($=J/2$) pulses are applied, with polarizations $\sigma^+$, $\sigma^-$, and in iteration (3), one ($=J/2^2$) $\sigma^+$ pulse is applied. This sequence can be followed by several cycles of conventional optical pumping---i.e., repetitions of iteration (3)---to further increase the end-state population.}
\end{figure}

We begin by considering the first scheme applied to a $J\to J'=J$ system with an initially unpolarized ground state. (A $J\to J'=J$ transition turns out to be more advantageous than $J\to J'=J\pm1$ transitions in terms of power requirements, as discussed in Sec.~\ref{sec:lightIntensity}.) In the initial stage, we apply a sequence of alternating $\sigma^+$ and $\sigma^-$ AFP pulses to ``fold'' the atomic populations in half, so that the atoms in the ground-state sublevels with $m<0$ are transferred to the upper-state sublevels with $m'>0$ [as shown in Fig.~\ref{PulseLevelDiagrams44}, iteration (1)]. This takes approximately $J$ pulses, which must be completed in a time shorter than the upper-state lifetime. We then wait for the upper-state atoms to decay. They decay into the already populated ground-state sublevels, approximately doubling the populations of the occupied levels. This process is then iterated, again folding the atomic populations in half---which requires approximately $J/2$ pulses for the second iteration---and allowing spontaneous decay to occur [Fig.~\ref{PulseLevelDiagrams44}, iteration (2)]. Approximately $\log_2 2J$ iterations are performed, with the $i$th iteration requiring about $J/2^{i-1}$ pulses. At this point, most of the atoms are in the end state, and additional cycles of conventional optical pumping can be performed to further increase the end-state population.

% \begin{figure}
% \includegraphics{PulseLevelDiagrams44a}
% \caption{\label{PulseLevelDiagrams44}Pulse sequence and resulting sublevel populations, obtained from the numerical model described in Sec.~\ref{sec:NumericalModel}, for the folding method applied to a $J=4\to J'=4$ system. The top diagram shows the initial state (unpolarized ground state and unpopulated upper state). Three iterations of the population folding procedure are applied, each followed by spontaneous decay of the upper state. In iteration (1), four ($=J$) AFP pulses are applied, with polarizations $\sigma^+$, $\sigma^-$, $\sigma^+$, $\sigma^-$. In iteration (2), two ($=J/2$) pulses are applied, with polarizations $\sigma^+$, $\sigma^-$, and in iteration (3), one ($=J/2^2$) $\sigma^+$ pulse is applied. This sequence can be followed by several cycles of conventional optical pumping---i.e., repetitions of iteration (3)---to further increase the end-state population.}
% \end{figure}

If atoms in the upper state decay only to the ground state under consideration, and not to any other metastable levels, this procedure can be used to transfer all of the atoms to the end state. If the branching ratio $R<1$, however, some atoms will be lost. In this case, we can estimate the final population $p_J^\text{folding}$ of the end state ($m=J$ sublevel) as follows. The initial population of the end state is $1/(2J+1)$. In each iteration of the folding procedure, the end-state population is multiplied by a factor of approximately $1+R$. Assuming $\log_2(2J+1)$ iterations, we have
\begin{equation}\label{eq:foldingestJJ}
	p_J^\text{folding} 
	\approx \frac{\br{1+R}^{\log_2(2J+1)}}{2J+1}
	= \br{2J+1}^{-1+\log_2\br{1+R}}. 
\end{equation}

On the other hand, in conventional optical pumping, atoms that begin in the ground-state sublevel $\ket{m}$ must undergo an average of about $J-m$ spontaneous decays before reaching the end state. Thus we can estimate the final population $p_J^\text{OP}$ for conventional optical pumping as
\begin{equation}\label{eq:tradestJJ}
	p_J^\text{OP} \approx \frac{1}{2J+1}\sum_{m=-J}^J R^{J-m}
	=\frac{1-R^{2J+1}}{\br{2J+1}\br{1-R}}.
\end{equation}
For large $J$ and $R<1$, the end-state population drops as $J^{-1}$, compared to the slower falloff $J^{-1+\log_2\br{1+R}}$ for the folding scheme. The ratio of the end-state populations for the two methods is then approximately
\begin{equation}
	\frac{p_J^\text{folding}}{p_J^\text{OP}} \approx \br{1-R^2}J^{\log_2(1+R)},
\end{equation}
where the exponent $\log_2(1+R)$ can vary between zero and one.

\begin{figure}
\includegraphics{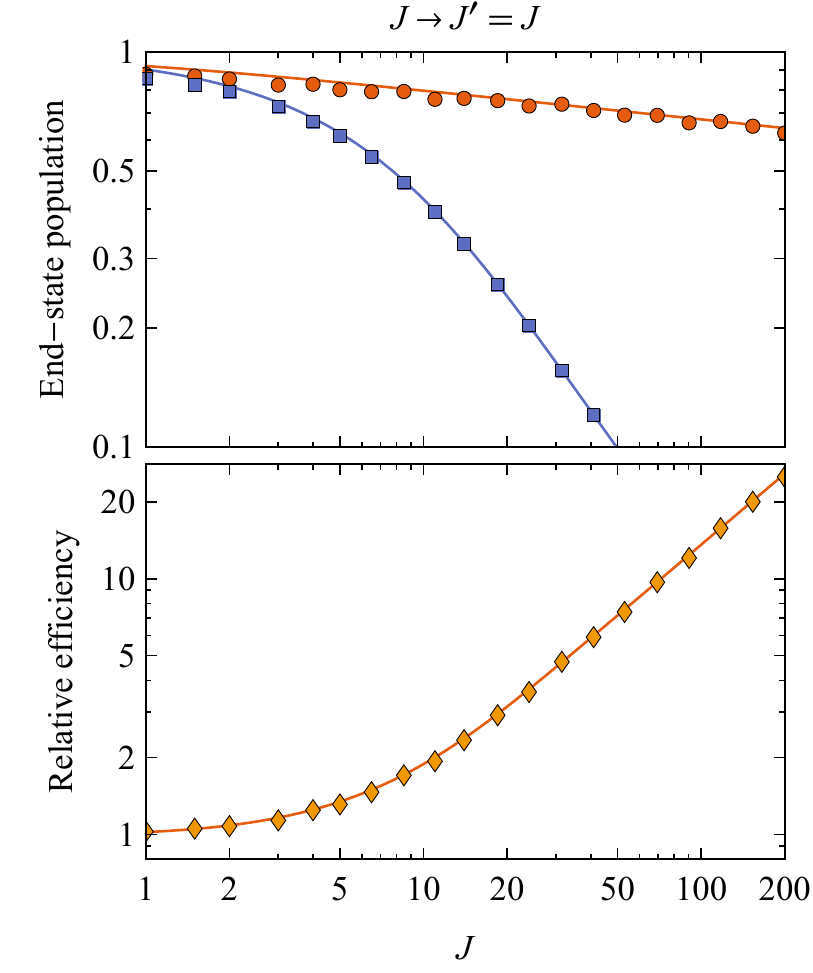}
\caption{\label{PopulationsJJ}(Top) The points show final end-state populations for the folding method (red circles) and for conventional optical pumping (blue squares) for $J\to J'=J$ transitions with various $J$ and $R=0.9$, calculated using the numerical model described in Sec.~\ref{sec:NumericalModel}. The curves show the estimates given in Eqs.~\eqref{eq:foldingestJJ} and \eqref{eq:tradestJJ}, respectively. (Bottom) The ratio of the calculated end-state populations for the two methods (orange diamonds), along with the curve showing the corresponding ratio of the estimates.}
\end{figure}

\begin{figure}
\includegraphics{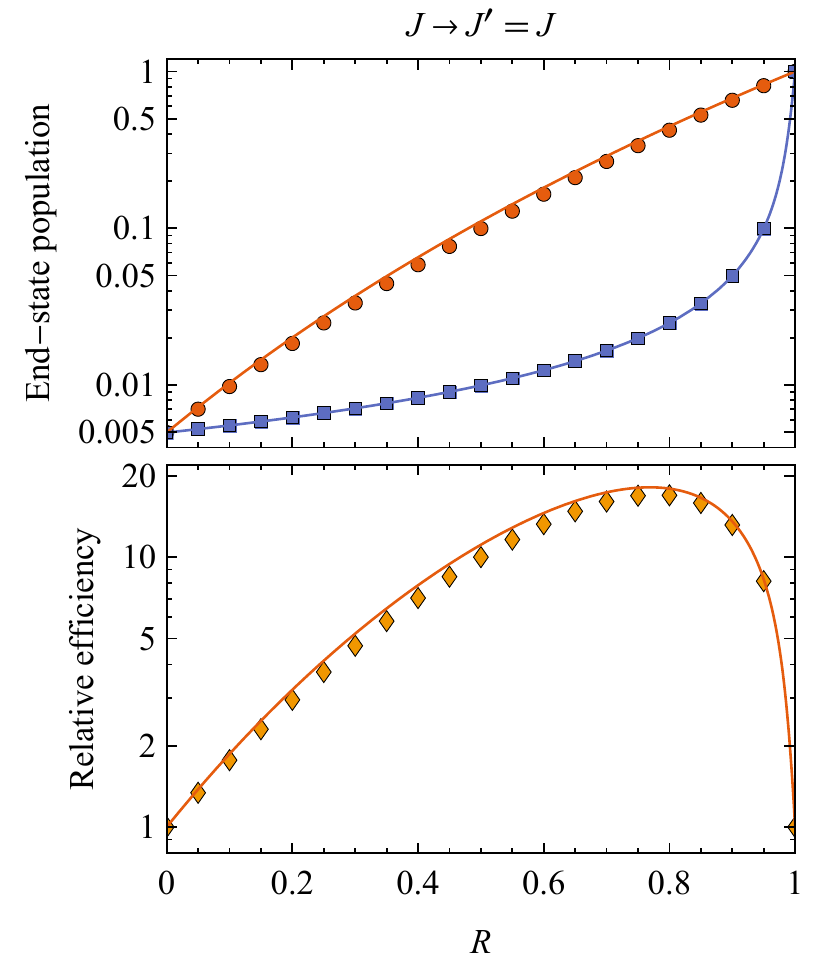}
\caption{\label{PopulationsRJJ}(Top) Final estimated (curves) and calculated (points) end-state populations for a $J\to J'=J$ transition for the folding method (red circles) and conventional optical pumping (blue squares) as in Fig.~\ref{PopulationsJJ}, here as a function of $R$ with $J=100$. (Bottom) The ratio of the end-state populations for the two methods (orange diamonds).}
\end{figure}

The estimates \eqref{eq:foldingestJJ} and \eqref{eq:tradestJJ} are plotted as curves in Figs.~\ref{PopulationsJJ} and \ref{PopulationsRJJ} as a function of $J$ and $R$, respectively, and compared to the results for particular values of $J$ obtained from the detailed numerical model described in Sec.~\ref{sec:NumericalModel}. Note that the estimates are defined mathematically for $J$ as a continuous parameter, although of course they only have physical meaning for integer and half-integer $J$.

\subsection{Folding method for $J\to J'=J-1$ systems}

\begin{figure}
\includegraphics{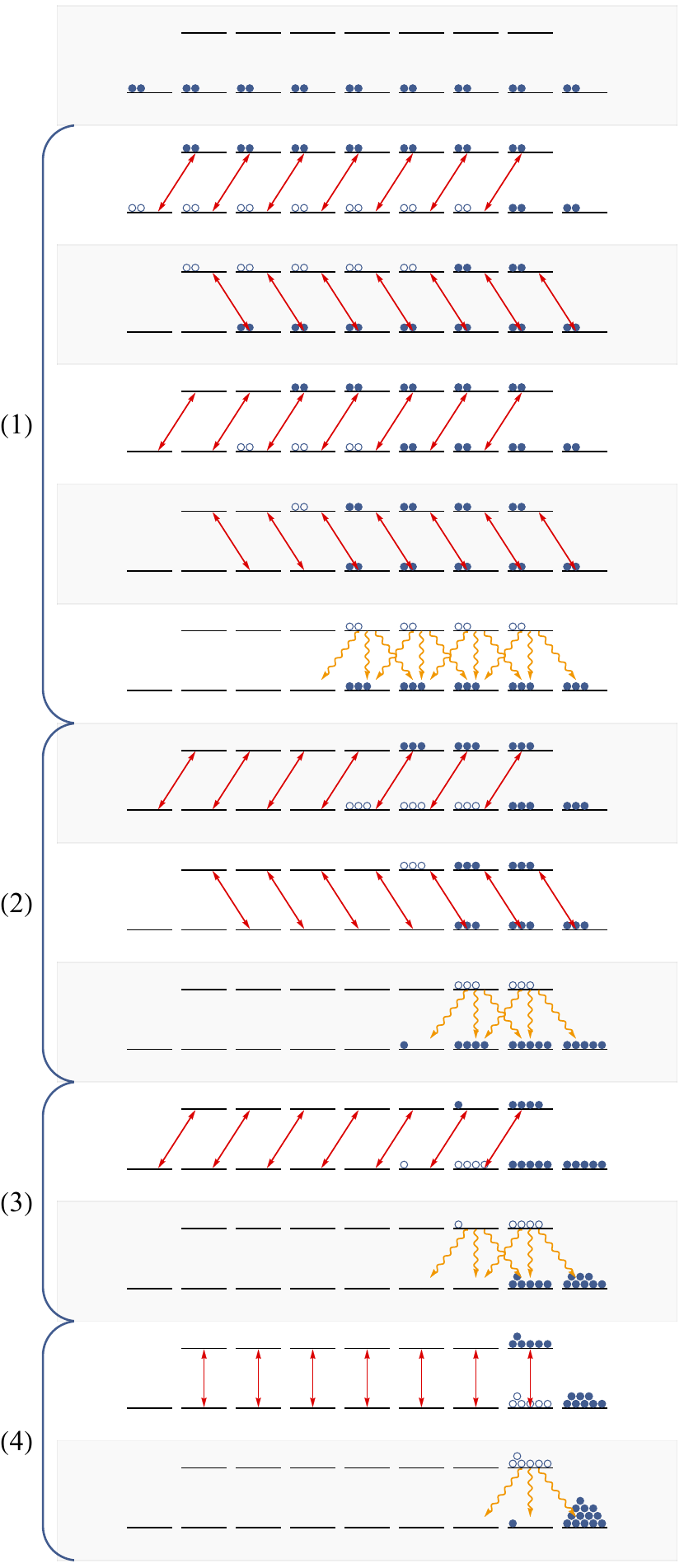}
\caption{\label{PulseLevelDiagrams43}Level diagrams showing the pulse sequence and resulting sublevel populations for the folding method applied to a $J=4\to J'=3$ system, as in Fig.~\ref{PulseLevelDiagrams44}. Three iterations of the population folding procedure are applied, each followed by spontaneous decay of the upper state; there is then a fourth iteration corresponding to conventional optical pumping. In iteration (1), four AFP pulses are applied, with polarizations $\sigma^+$, $\sigma^-$, $\sigma^+$, $\sigma^-$. In iteration (2), two pulses are applied, with polarizations $\sigma^+$, $\sigma^-$. In iteration (3), a $\sigma^+$ pulse is applied. Iteration (4) corresponds to a conventional optical pumping cycle with $\pi$-polarized light.}
\end{figure}

The folding method works much the same in $J\to J'=J-1$ systems (Fig.~\ref{PulseLevelDiagrams43}) as it does in $J\to J'=J$ systems. The main difference is that after the folding iterations are complete, the atoms are split between the ground-state $\ket{m=J}$ and $\ket{m=J-1}$ sublevels, as both of these are dark states for $\sigma^+$ light [Fig.~\ref{PulseLevelDiagrams43}, iteration (3)]. One or two additional optical-pumping cycles with $\pi$-polarized light will transfer most of the population of $\ket{m=J-1}$ to the end state [Fig.~\ref{PulseLevelDiagrams43}, iteration (4)]. This optical-pumping stage is quite efficient, especially for large $J$, because for $J\to J'=J-1$ transitions atoms tend to spontaneously decay to sublevels for which the magnitude of $m$ is largest; i.e., $\ket{m'}\to\ket{m=m'+1}$ decays are strongly favored for positive $m$. Thus atoms excited from $\ket{m=J-1}$ to $\ket{m'=J-1}$ by $\pi$-polarized light tend to fall immediately into $\ket{m=J}$ [Fig.~\ref{PulseLevelDiagrams43}, iteration (4)]. For low $J$, cycles of optical pumping with $\sigma^+$-polarized light can also be used to transfer any residual population of $\ket{m=J-2}$ to the end state.

The end-state population can be estimated to be approximately the same as in the $J\to J'=J$ case. This estimate is somewhat less accurate for small $J$, where the $\ket{m'=J-1}\to\ket{m=J}$ decay is not as strongly favored. 

In contrast, conventional optical pumping is actually more efficient in $J\to J'=J-1$ systems than in $J\to J'=J$ systems, because of the tendency for decaying atoms to fall away from $m=0$. This means that for large $J$, atoms initially in ground-state sublevels with $m\ge0$ tend to have their $m$ increased by two in each cycle of conventional optical pumping with $\sigma^+$ light. Thus, in multiple cycles of optical pumping, these atoms will  undergo only about $(J-m)/2$ spontaneous decays before reaching the end state. We actually get a somewhat more accurate estimate by assuming a slightly larger number, $(J-m)/1.8$. As a rough estimate we write
\begin{equation}\label{eq:tradestJJm}
\begin{split}
	p_J^\text{OP} 
	&\approx 
	\frac{1}{2J+1}\br{\sum_{m=-J}^{-1} R^{J-m} + \sum_{m=0}^{J} R^{\br{J-m}/1.8}}\\
	&=\frac{1}{2J+1}\br{
	   \frac{R^{J+1}\br{1-R^J}}{\br{1-R}}
	   +\frac{1-R^{\br{J+1}/1.8}}{1-R^{1/1.8}}
	  }.
\end{split}
\end{equation}
Here we have assumed that atoms initially in ground-state sublevels with $m<0$ undergo an average of $J-m$ spontaneous decays before reaching $\ket{m=J}$ [first term in the square brackets of Eq.~\eqref{eq:tradestJJm}]. The actual dependence on $m$ for $m<0$ is more complicated than this, but this estimate is reasonably accurate for low $J$, where decays away from $m=0$ are not so strongly favored. It is not accurate for large $J$, but in that case, with $R<1$, it is the second term in the square brackets of Eq.~\eqref{eq:tradestJJm} that dominates, so the accuracy of the first term is not important.

\begin{figure}
\includegraphics{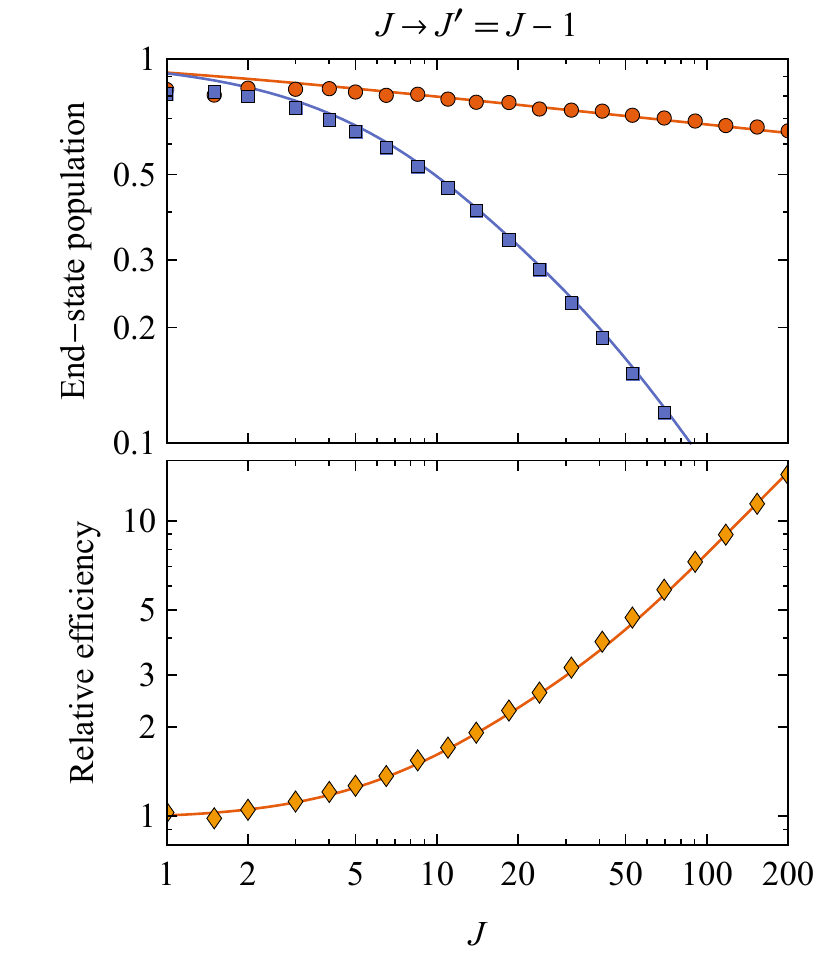}
\caption{\label{PopulationsJJm}(Top) Comparison of final end-state populations for the folding method (red circles) and for conventional optical pumping (blue squares) as in Fig.~\ref{PopulationsJJ} but for $J\to J'=J-1$ transitions with $R=0.9$. The curves show the estimates given in Eqs.~\eqref{eq:foldingestJJ} and \eqref{eq:tradestJJm}, respectively. (Bottom) The ratio of the calculated end-state populations for the two methods (orange diamonds), along with the corresponding ratio of the estimates.}
\end{figure}

\begin{figure}
\includegraphics{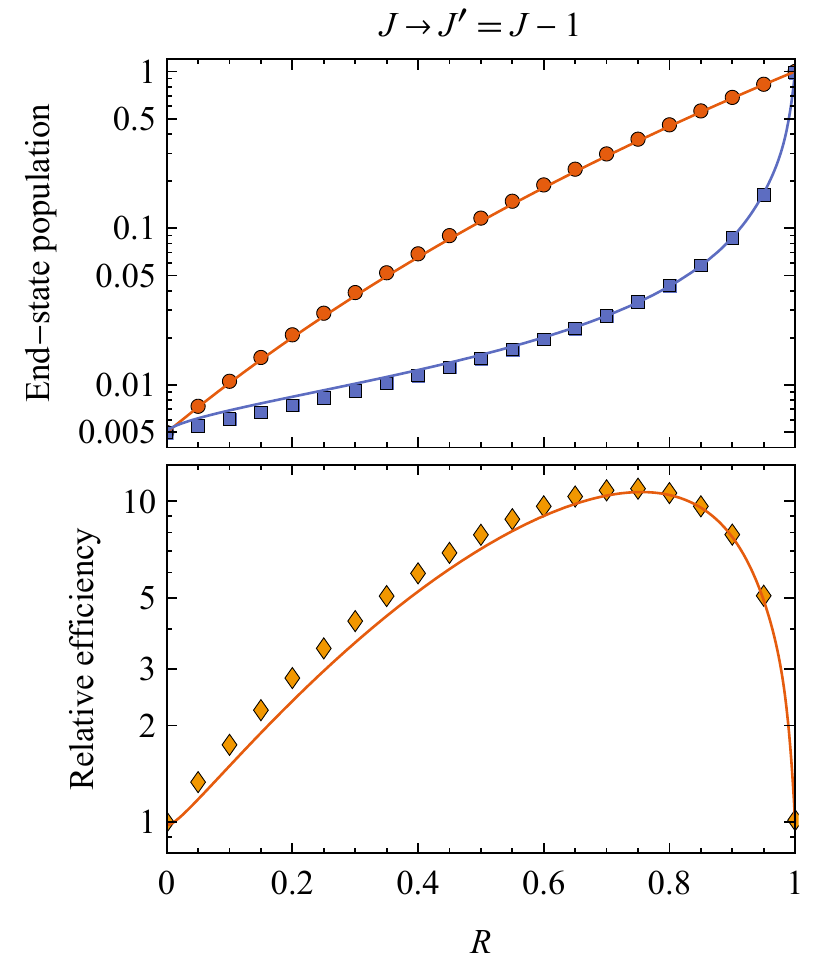}
\caption{\label{PopulationsRJJm}(Top) Final estimated (curves) and calculated (points) end-state populations for a $J\to J'=J-1$ transition with $J=100$ for the folding method (red circles) and conventional optical pumping (blue squares) as in Fig.~\ref{PopulationsJJm}, here as a function of $R$. (Bottom) The ratio of the end-state populations for the two methods (orange diamonds).}
\end{figure}

The estimates \eqref{eq:foldingestJJ} and \eqref{eq:tradestJJm} are plotted in Figs.~\ref{PopulationsJJm} and \ref{PopulationsRJJm}, and compared to the values obtained from the numerical model described in Sec.~\ref{sec:NumericalModel}. 

\subsection{Folding method for $J\to J'=J+1$ systems}

\begin{figure}
\includegraphics{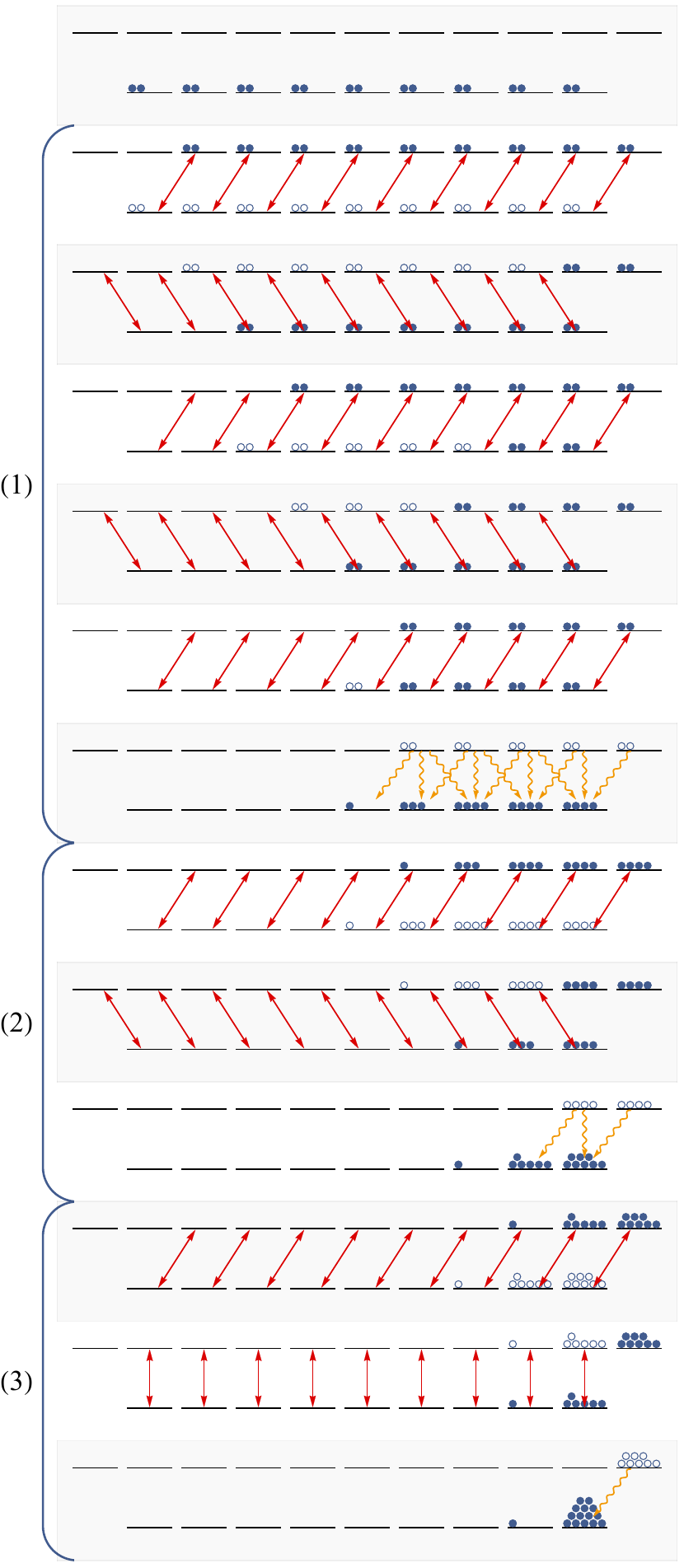}
\caption{\label{PulseLevelDiagrams45}Level diagrams showing the pulse sequence and resulting sublevel populations for the folding method applied to a $J=4\to J'=5$ system, as in Fig.~\ref{PulseLevelDiagrams44}. Three iterations of the population folding procedure are applied, each followed by spontaneous decay of the upper state. In iteration (1), five AFP pulses are applied, with polarizations $\sigma^+$, $\sigma^-$, $\sigma^+$, $\sigma^-$, $\sigma^+$. In iteration (2), two pulses are applied, with polarizations $\sigma^+$, $\sigma^-$. In iteration (3), a $\sigma^+$ pulse and then a $\pi$-polarized pulse are applied.}
\end{figure}

For $J\to J'=J+1$ systems (Fig.~\ref{PulseLevelDiagrams45}), there are no dark states in the ground state. In addition, upper-state atoms tend to decay \emph{toward} the $m=0$ ground-state sublevel in such systems (i.e., $\ket{m'}\to\ket{m=m'-1}$ transitions are favored for positive $m'$). This means that, as for the $J\to J'=J-1$ case, after application of the folding procedure atoms tend to end up split between the $\ket{m=J-1}$ and $\ket{m=J}$ ground-state sublevels [Fig.~\ref{PulseLevelDiagrams45}, iteration (2)]. For $R<1$, optical pumping will be ineffective in combining the atoms in $\ket{m=J}$. However, a $\sigma^+$ AFP pulse followed by a $\pi$-polarized AFP pulse will excite the $\ket{m=J}$ atoms to $\ket{m'=J+1}$ while coherently transferring the atoms from $\ket{m=J-1}$ to $\ket{m=J}$---subsequent spontaneous decay will then combine the atoms in $\ket{m=J}$ [Fig.~\ref{PulseLevelDiagrams45}, iteration (3)].

With the above modification, the folding procedure can be estimated to produce approximately the same final end-state population as for the $J\to J'=J$ and $J\to J'=J-1$ cases.

On the other hand, conventional optical pumping is ineffective for $J\to J'=J+1$ systems because of the characteristics cited above. For low $J$ and not-too-small $R$, optical pumping can at least be used to deplete the ground-state sublevels with $m<J$, leaving a population of approximately $1/(2J+1)$ in the end state. For larger $J$, however, $\ket{m=J}$ is depleted nearly as fast as the other sublevels. Therefore, we do not compare the efficiency of the folding procedure to that of conventional optical pumping for the $J\to J'=J+1$ case.

\begin{figure}
\includegraphics{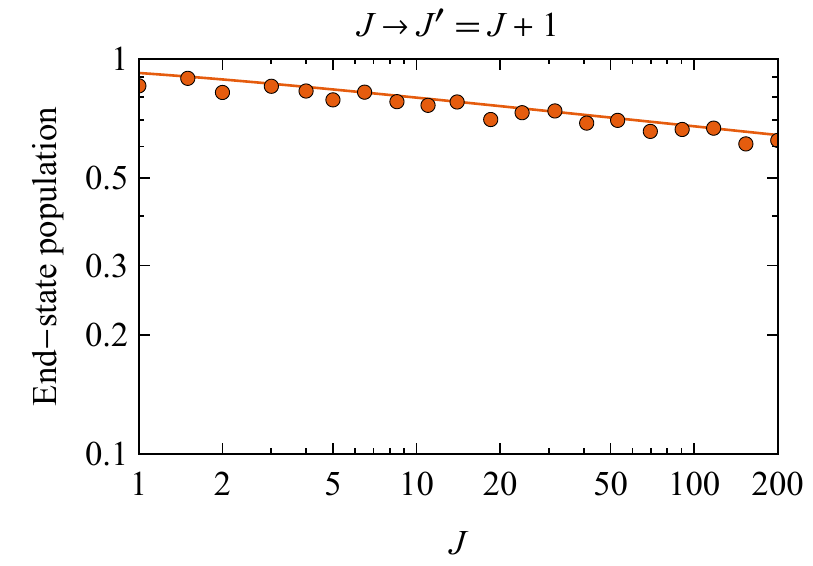}
\caption{\label{PopulationsJJp}Final end-state populations for the folding method (red circles) for $J\to J'=J+1$ transitions with $R=0.9$. The curve shows the estimate given in Eq.~\eqref{eq:foldingestJJ}.}
\end{figure}

\begin{figure}
\includegraphics{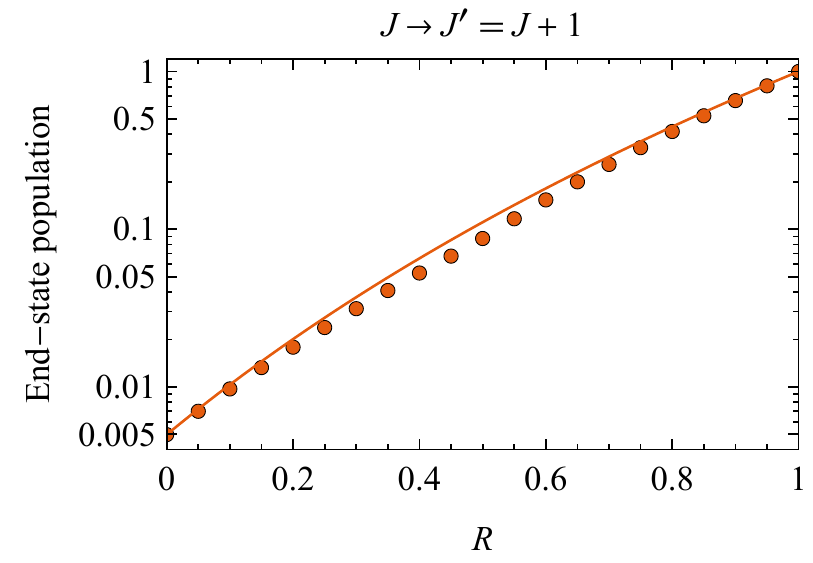}
\caption{\label{PopulationsRJJp}Final estimated and calculated end-state populations produced by the folding method for a $J\to J'=J+1$ transition with $J=100$ as a function of $R$.}
\end{figure}

The estimate \eqref{eq:foldingestJJ} is compared to the results obtained from the numerical model in Figs.~\ref{PopulationsJJp} and \ref{PopulationsRJJp}.  

\subsection{Folding method with a shelving state}
\label{sec:FoldingShelving}

\begin{figure}
\includegraphics{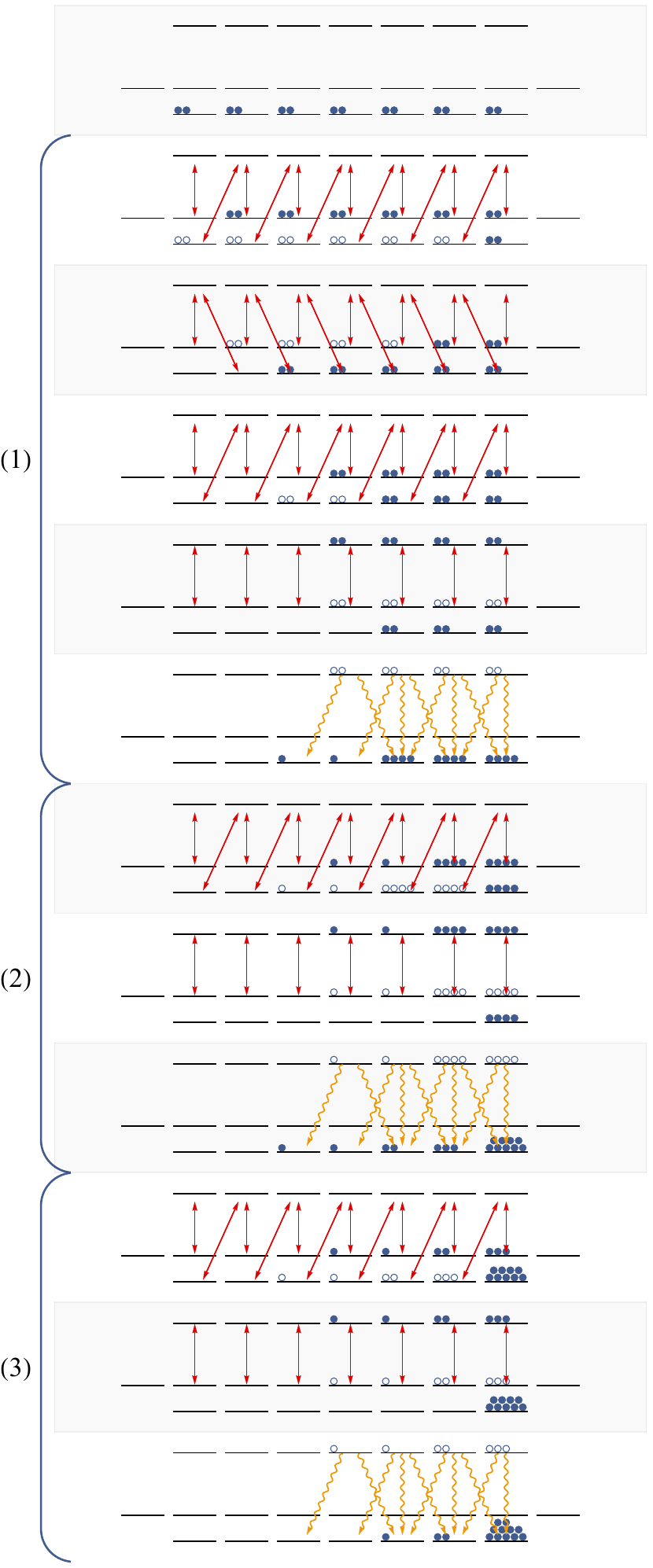}
\caption{\label{PulseLevelDiagramsFoldingShelving334}Pulse sequence and resulting sublevel populations for the folding method applied to a $J=3\to J'=3$ system with an additional $J=4$ shelving state. Three iterations of the population folding procedure are applied, each followed by excitation from the shelving state to the excited state with $\pi$-polarized light and subsequent spontaneous decay to the ground state. In iteration (1), three two-photon AFP pulses are applied, and in iterations (2) and (3), one two-photon AFP pulse is applied.}
\end{figure}

The folding method can also be implemented using a metastable shelving state, so that atoms do not have to be stored in the excited state during the coherent-transfer steps. This can be advantageous when the excited-state lifetime is short. A two-step coherent transfer method is used to exchange populations between the ground state and the shelving state, as shown in Fig.~\ref{PulseLevelDiagramsFoldingShelving334} for the example case of a $J=3$ ground state, $J'=3$ excited state, and $J=4$ shelving state. (As for the case without a shelving state, the power requirements are reduced when the ground and excited states have the same $J$, as will be shown in Sec.~\ref{sec:lightIntensity}.) It is assumed that the branching ratio of the excited state to the ground state is larger than that to the shelving state. This is necessary to avoid additional cycles of conventional optical pumping. For the folding scheme to work as we have described, the coherent-transfer method must swap the populations of the ground and shelving states when both are initially populated. This rules out conventional STIRAP as a transfer mechanism, since it works in one direction only; atoms already in the target state will be transferred to the intermediate state, whence they may be lost to spontaneous decay. Two-photon AFP can be used instead, or a bi-directional variant of STIRAP~\cite{Mukherjee2011}. Another requirement on the coherent-transfer method is that it not leave any atoms in the excited state, even when an excited-state sublevel is a dark state for one of the light polarizations. An example of this is the $m'=3$ excited-state sublevel in the second diagram of iteration (1) of Fig.~\ref{PulseLevelDiagramsFoldingShelving334}. There, the $\pi$-polarized light would transfer atoms from the $m=3$ shelving-state sublevel to the $m'=3$ sublevel unless the one-photon detuning is sufficiently large. (In the figure, arrows are shorter to indicate the one-photon detuning.) The necessity to detune the light fields can increase the light power requirements for the method, as discussed in Sec.~\ref{sec:lightIntensity}.

In the pulse sequence shown in Fig.~\ref{PulseLevelDiagramsFoldingShelving334}, three iterations of the population folding procedure are applied. Each iteration consists of a series of two-photon AFP pulses, with $\pi$-polarized light detuned from the transition between the shelving state and the excited state, and alternating $\sigma^+$ and $\sigma^-$ polarized light detuned from the transition between the ground state and the excited state. In iteration (1), three two-photon AFP pulses are applied, and in iterations (2) and (3), one two-photon AFP pulse is applied. Each iteration is followed by excitation from the shelving state to the excited state with $\pi$-polarized light and subsequent spontaneous decay to the ground state. After this pulse sequence is completed, additional cycles of conventional optical pumping could be employed to transfer the residual population of the $m<J$ sublevels to the end state.

\subsection{Maximally efficient method with two shelving states}

\begin{figure}
\includegraphics{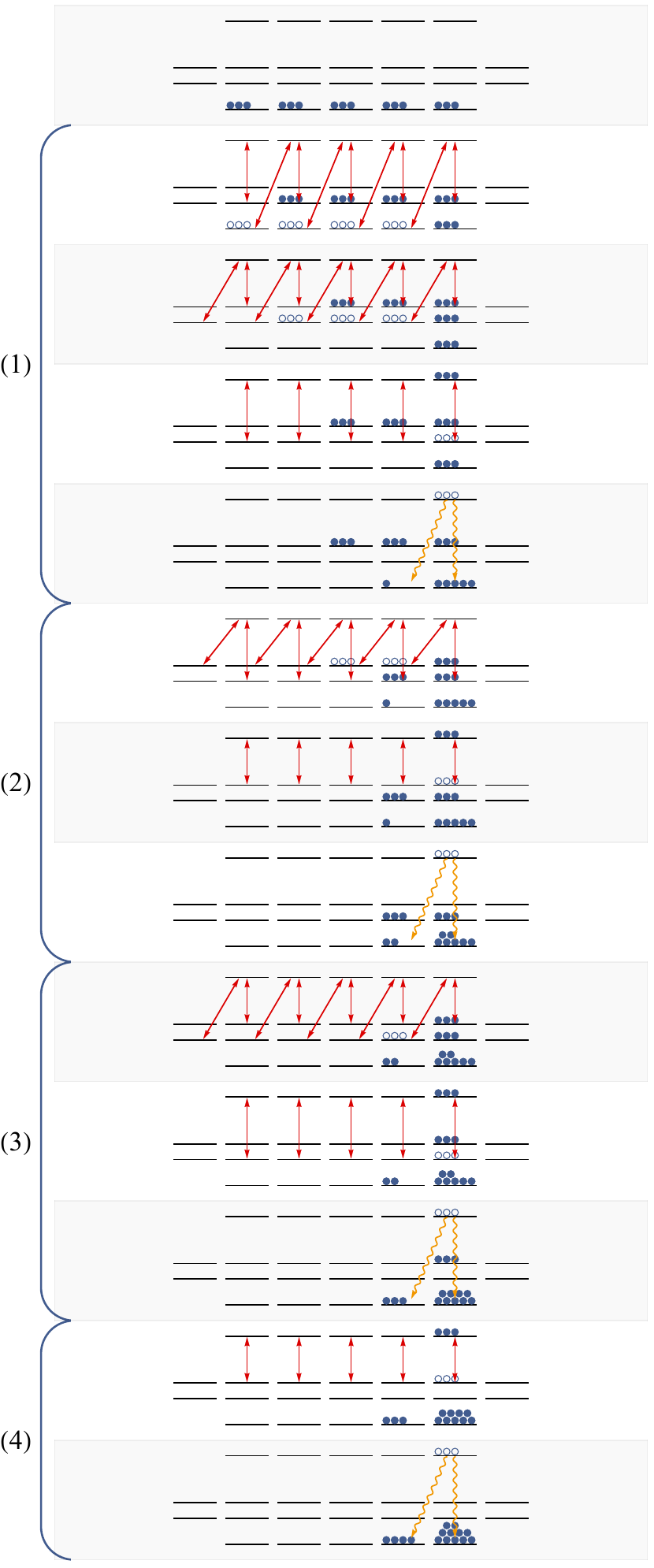}
\caption{\label{PulseLevelDiagramsTwoShelves2233}Pulse sequence for the two-shelving-state method described in the text applied to a $J=2\to J'=2$ system with two $J=3$ shelving states. Here we neglect spontaneous decay into the shelving states compared to the decay channel to the ground state. Four iterations are shown; an additional cycle of conventional optical pumping (not shown) can be used to transfer the residual population into the end state.}
\end{figure}

Even more efficient optical pumping can be achieved with a scheme that uses \emph{two} shelving states, diagrammed in Fig.~\ref{PulseLevelDiagramsTwoShelves2233} for a system with a $J=2$ ground state, $J'=2$ excited state, and two $J=3$ shelving states. The atoms are initially transferred from the ground state to one of the shelving states using STIRAP with $\sigma^+$- and $\pi$-polarized light fields, leaving the end-state sublevel of the ground state occupied [Fig.~\ref{PulseLevelDiagramsTwoShelves2233}, first pulse of iteration (1)]. A similar STIRAP pulse is then used to transfer the atoms to the other shelving state, leaving one sublevel of the first shelving state occupied---$\pi$-polarized light is then used to excite this population to the upper state, whence it decays to the ground state. In subsequent iterations, the atoms are transferred back and forth between the shelving states, each time leaving atoms in one sublevel, which are then transferred to the ground state via excitation and decay. After the final iteration---iteration (4), in this case---an additional cycle of conventional optical pumping (not shown in the figure) can be used to transfer the population of the $m=J-1$ sublevel to the end state.

In all, there are $2J+1$ spontaneous-decay cycles, each with about $1/(2J+1)$ of the atoms participating, so each atom undergoes an average of about one spontaneous decay in creating the stretched state. As discussed in Sec.~\ref{sec:intro}, this is the fewest allowed by the second law of thermodynamics. For large $J$ and branching ratio $R>0$, the final end-state population is approximately $R$, independent of $J$. Thus, from Eq.~\eqref{eq:tradestJJ}, the efficiency relative to conventional optical pumping for large $J$ and $0<R<1$ is about $2JR(1-R)$. The largest efficiency gain, a factor of $J/2$, occurs for $R=1/2$.

In this method, as opposed to the folding method with a shelving state, only one-way coherent transfer is required, so conventional STIRAP can be used. Another attractive feature of this method is that only one circular light polarization is used, which simplifies the implementation and also avoids the possibility for atoms to be trapped in excited-state dark states, so the STIRAP fields can be tuned on one-photon resonance.

The main drawback to this method is that the $2J+1$ spontaneous-decay cycles occur sequentially, increasing the time required for the complete optical-pumping sequence compared to the folding method. This may cause difficulty for experimental implementation if the upper-state lifetime is long.

\section{Numerical model}
\label{sec:NumericalModel}

To verify the accuracy of the estimates for the folding method with no shelving states made in Sec.~\ref{sec:description}, we can calculate the effect of the coherent population transfer and spontaneous decay for a particular $J\to J'$ transition and a particular realization of the pulse sequence. We can assume that, at the beginning and end of each AFP pulse, the atomic state does not possess any coherences. Thus we can represent the atomic state as a column vector $b$ of populations of the sublevels, and the effect of each AFP pulse and spontaneous decay event by matrix multiplication. The initial population vector $b^0$ is given by $b^0_m=1/(2J+1)$ for ground-state sublevels $m$, and $b^0_{m'}=0$ for upper-state sublevels $m'$. 

The $\sigma^\pm$ (and $\pi$) AFP pulses are represented by unitary matrices $U_\pm$ ($U_0$) that swap the populations of the ground-state sublevels $\ket{m}$ with those of the excited-state sublevels $\ket{m'=m\pm1}$ ($\ket{m'=m}$). The effect of spontaneous emission is described by a matrix $S$ whose matrix elements are all zero except for $S_{mm'}$ and $S_{mm}$, where $m$ refers to a ground-state sublevel and $m'$ refers to an upper-state sublevel. The matrix elements
\begin{equation}
S_{mm'} = R (2J'+1) 
\left( \begin{array}{ccc} 
J & 1 & J'\\
-m & m-m' & m'
\end{array}\right)^2
\end{equation}
describe the transfer of atoms from the upper state to the ground state, and are given by the corresponding decay rate for each transition as a fraction of the total decay rate of the upper state (this is obtained from the Wigner-Eckart theorem, see Ref.~\cite{Auzinsh2010}, Sec.~12.1, for a derivation). Here
\begin{equation*}
\left( \begin{array}{ccc} \cdot & \cdot & \cdot\\
\cdot & \cdot & \cdot
\end{array}\right)
\end{equation*}
is the 3-j symbol. The matrix elements 
\begin{equation}
S_{mm} = 1
\end{equation}
specify that atoms initially in the ground state stay in the ground state.

Each iteration of length $k$ consists of left-multiplication by a series of $k$ alternating $U_+$ and $U_-$ matrices, followed by an $S$ matrix. For example, if the first iteration is length three, the population after the first iteration is given by $b^1=SU_+U_-U_+b^0$. The standard pulse sequence for a $J\to J'=J$ system consists of $n=\lceil\log_22J\rceil$ iterations, with the length of the $i$th iteration given by $k_i=\lceil J/2^{i-1}\rceil$. Here $\lceil\,\cdot\,\rceil$ indicates the ceiling function, which gives the smallest integer greater than or equal to its argument. The population of the end state is then given by $b^{n}_J$. The $n$ iterations are followed by a number of conventional optical pumping cycles, simulated by multiplication with alternating $U_+$ and $S$ matrices, to transfer a fraction of the residual population in sublevels with $m<J$ into the end state.

For a $J\to J'=J-1$ system the standard pulse sequence is the same, except that it is followed by optical pumping with a combination of $\sigma^+$ and $\pi$-polarized light, simulated by repeated multiplication by the sequence $SU_+SU_0$.

For a $J\to J'=J+1$ system the standard pulse sequence is $n=\lceil\log_22J\rceil-1$ iterations, with the length of the $i$th iteration given by $k_i=\lceil J/2^{i-1}\rceil+1$, followed by a final iteration described by multiplication by $SU_0U_+$, as shown in Fig.~\ref{PulseLevelDiagrams45}. For small $R$, it may also be advantageous to use a variant of the standard sequence: $n=\lceil\log_22J\rceil-2$ iterations each of length $k_i=\lceil J/2^{i-1}\rceil$, followed by a final iteration described by $SU_0U_+$.

For a given system, the pulse sequence can be numerically optimized by varying the iteration lengths $k_i$, either with the same variation for all $k_i$, or with individual variations for each. To minimize the computational effort in the latter case, we can restrict the the variation of each $k_i$ to one greater or less than the standard sequence length. Calculating $b^{n}_J$ for all combinations of lengths $k'_i=k_i-1$, $k_i$, $k_i+1$ requires multiplying $3^n$ sequences of matrices, which is still tractable for $J$ of the order of a few hundred, so that $n$ is less than ten.
% \SR{I have only done the simple variation of all iteration lengths together here. Maybe Konrad wants to generate the full optimized results suitable for plotting in the format shown here?}

Conventional optical pumping is simulated by applying multiple iterations of the form $SU_+$ until the ground-state populations stabilize. For a $J\to J'=J$ system, the remaining atoms are left in the $\ket{m=J}$ sublevel. For a $J\to J'=J-1$ system, atoms are left in the $\ket{m=J}$ and $\ket{m=J-1}$ sublevels. In this case, additional optical pumping with $\sigma^+$ and $\pi$-polarized light is then simulated by applying multiple iterations of the form $SU_+SU_0$. As stated above, conventional optical pumping is ineffective for $J\to J'=J+1$ systems, so we have not performed calculations for that case.

The results of the numerical simulation are compared to the efficiency estimates in Figs.~\ref{PopulationsJJ}, \ref{PopulationsRJJ}, \ref{PopulationsJJm}, \ref{PopulationsRJJm}, \ref{PopulationsJJp}, and \ref{PopulationsRJJp}.

\section{Light intensity requirements}
\label{sec:lightIntensity}

\subsection{Adiabatic condition}

Proper implementation of STIRAP with resonant light fields requires the satisfaction of the adiabatic condition~\cite{Bergmann1998}
\begin{equation}\label{eq:adiabatic}
	\tau_p\Omega_R \gg 1,
\end{equation}
where $\tau_p$ is the overlap time between the pump and Stokes pulses and $\Omega_R$ is the quadrature sum of the Rabi frequencies of the two pulses. A general adiabatic condition for AFP~\cite{Nguyen2000Efficient} also reduces to Eq.~\eqref{eq:adiabatic} for some parameter ranges. For AFP there is one frequency-swept pulse; then $\tau_p$ corresponds to the time taken for the light frequency to sweep through resonance, and $\Omega_R$ is the Rabi frequency induced by the pulse. In the case of AFP, $\tau_p$ must be shorter than the upper-state lifetime, in the case of STIRAP the only limitation is the (potentially much longer) light--atom interaction time, given, for example, by the transit time of atoms through the laser beam. The upper limit to $\tau_p$ imposes a minimum requirement on the laser intensity, which depends on the specific sublevels $m\to m'$ involved in the transition---the condition must be satisfied for all of the $m\to m'$ transitions for the proposed method to be effective.

The Rabi frequency is written in terms of the electric dipole matrix element $\bra{J'm'}d_q\ket{Jm}$ and optical electric field amplitude $E$ as 
\begin{equation}
\Omega_R=\mel{J'm'}{d}{Jm}E/\hbar,
\end{equation}
where $\hbar$ is Planck's constant. The required light intensity is given in terms of $E$ by $I=c\epsilon_0E^2/2$, where $\epsilon_0$ is the electric constant. The angular part of the dipole matrix element can be factored out using the Wigner--Eckart theorem:
\begin{equation}
\bra{J'm'}d_q\ket{Jm}^2
=\langle J'\|d\|J\rangle^2\left(\begin{array}{ccc}
J'  & 1 & J \\
-m' & q & m 
 \end{array} \right)^2,
\end{equation}
where $d_q$ is the spherical component of the dipole matrix element corresponding to the light polarization in question ($q=\pm1, 0$ for $\sigma^\pm, \pi$), and the reduced dipole matrix element $\langle J'\|d\|J\rangle$ is independent of $m$, $m'$, and $q$. From the above, we have for the required light intensity
\begin{equation}\label{eq:adiabaticI}
\begin{split}
	I %&=c\epsilon_0E^2/2 \\
%	&=c \epsilon_0 \hbar^2\Omega_R^2/\mel{Jm}{d}{J'm'}^2/2 \\
%	&=c \epsilon_0 \hbar^2\Omega_R^2/\mel{Jm}{d}{J'm'}^2/2 \\
%	&=\frac{c\epsilon_0\hbar^2\Omega_R^2}{2\langle J'\|d\|J\rangle^2}
%		\mqty(J'&1&J\\-m'&q&m)^{-2}\\
	&\gg\frac{c\epsilon_0\hbar^2}{2\tau_p^2\langle J'\|d\|J\rangle^2}
		\mqty(J'&1&J\\-m'&q&m)^{-2}.
\end{split}
\end{equation}

The angular factor can lead to a strong suppression of some $m\to m'$ transitions. For example, for a $J\to J'=J$ transition (with integer $J$) and $\sigma^+$ light, the $m=0\to m'=1$ transition is nearly unsuppressed, but the $m=J-1\to m'=J$ transition strength is suppressed by a factor
\begin{equation}
	\frac{\mel{J'J}{d_1}{J,J-1}^2}{\mel{J'1}{d_1}{J0}^2}
	=\frac{2}{J+1}.
\end{equation}
An even larger suppression is seen for $J\to J'=J+1$ transitions; there the $m=J\to m'=J+1$ transition is unsuppressed, while the $m=-J\to m'=-J+1$ transition strength is smaller by a factor
\begin{equation}
	\frac{\mel{J',-J+1}{d_1}{J,-J}^2}{\mel{J',J+1}{d_1}{JJ}^2}
	=\frac{1}{(J+1)(2J+1)}.
\end{equation}
Because the adiabatic criterion must be satisfied for all of the $m\to m'$ transitions, for a $J\to J'=J+1$ transition the light intensity must be a factor of $2J+1$ higher than for a $J\to J'=J$ transition with the same $J$ and reduced dipole matrix element. A similar result holds for $J\to J'=J-1$ transitions. Thus $J\to J'=J$ transitions may be preferred for practical realization. 

For the folding method with a shelving state, the one-photon detunings for the two fields used for coherent transfer must be large enough that atoms are not excited to the upper state even if only one field is applied, as discussed in Sec.~\ref{sec:FoldingShelving}. For short Gaussian pulses with $1/\tau_p\gg\Gamma$ this requires $\Delta\gg1/\tau_p$, as power-broadening is minimal in this situation. For large one-photon detuning, the adiabatic condition \eqref{eq:adiabatic} for coherent transfer is modified, to become~\cite{Gaubatz1990}
\begin{equation}\label{eq:detunedAdiabaticCondition}
	\frac{\tau_p\Omega_R^2}{\Delta} \gg 1.
\end{equation}
Combining this condition with $\Delta\gg1/\tau_p$, we have
\begin{equation}
	\tau_p\Omega_R \gg \sqrt{\tau_p\Delta} \gg 1,
\end{equation}
leading to a more stringent condition on the light intensity than for the resonant case.

\subsection{Optical pumping of K$_2$}

\begin{figure}[h]
\includegraphics{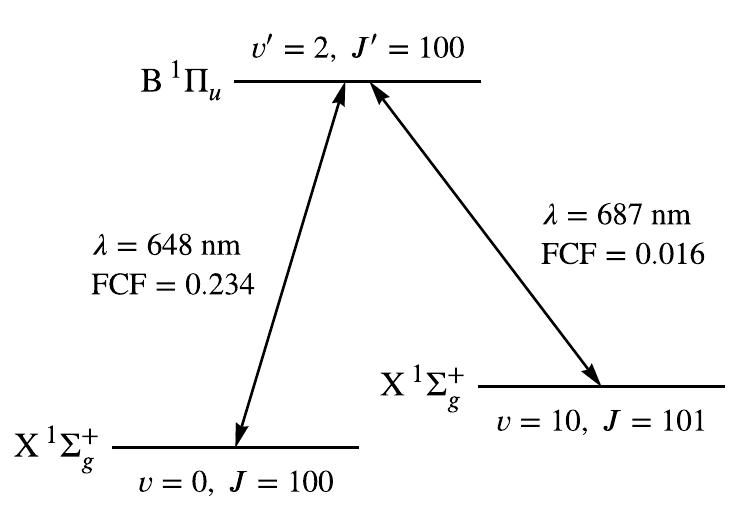}
\caption{\label{levels}Diagram of the levels of the K$_2$ molecule under consideration.}
\end{figure}

As an example we consider the K$_2$ molecule, which has transitions between the ground state, denoted by $\text{X}\,^1\Sigma_g^+$, and an excited state denoted by $\text{B}\,^1\Pi_u^+$. In particular, the ground-state rotational-vibrational level $v=0$,~$J=100$ has a strong transition to $\text{B}\,^1\Pi_u^+$ $v'=2$,~$J'=100$ (Fig.~\ref{levels}), with Franck--Condon factor $f_{FC} \approx 0.234$~\cite{FeberPriv} and wavelength $\lambda = 648$~nm (molecular constants were taken from~\cite{Huber1979}). A second, somewhat weaker (Franck--Condon factor $f_{FC} \approx 0.016$~\cite{FeberPriv}) transition couples the excited state rotational--vibrational level $v'=2,~J'=100$ to the ground-state level $v=10$,~$J=101$ with transition wavelength $\lambda = 687$~nm~\cite{Huber1979}.

% The vibrational level with $v=10$ is roughly\SR{I get $\rm (647~nm)^{-1}-(689~nm)^{-1}=942~cm^{-1}$.} 877~cm$^{-1}$ above the lowest vibrational level $v=0$.  If the K$_2$ sample is at room temperature, $300$~K, the population of the $v=10$ level will be around\SR{I get $\exp\br{-\hbar c\br{942~\rm cm^{-1}}/\br{k_B \br{\rm 300~K}}}=49\%$.} $1.5\%$ of that of $v=0$. Thus the $v=10$ level can function as a shelving state for the proposed folding method using STIRAP.\SR{If my numbers are correct, this won't work, because the shelving state is already occupied. Maybe we need to discuss cooling after all?}
%
% If, alternatively, we assume that we have a molecular beam, then the molecules are vibrationally cooled and we can assume that the state with $v=10$ is not populated at all. Depending on the beam expansion conditions, the population distribution among the ground state vibrational levels can differ. In a supersonic beam almost all the population will be in the lowest vibration level with $v=0$; in a slower beam some of the lowest vibrational levels will be populated, but the vibrational level populations will drop much faster than in the cell with the same temperature~\cite{??}.\SR{Cite Klass Bergmann's early papers.}

The dipole moment for these transitions is large: the reduced matrix element can be estimated as
\begin{equation}
 \| d \|  = \sqrt{f_{FC}\tilde{f}_{HL}\frac{1}{\tau}\frac{(2 J' + 1)3\pi\epsilon_0\hbar c^3}{\omega_{0}^{3}}},
 \label{eq1}
\end{equation}
where $\tau$ is the lifetime of the upper state, $\omega_{0}$ is the transition frequency, and the coefficient $\tilde{f}_{HL}$ is the reduced H\"{o}nl--London factor. For the weaker transition from $v=10$,~$J=102$ to $v'=2$,~$J'=101$, we take $\tau = 12$ ns~\cite{Ferber1979}, and $f_{FC}\approx0.016$, as mentioned above. The reduced H\"{o}nl--London factors for perpendicular $P$ ($J\to J'=J$) and $R$ ($J\to J'=J+1$) transitions with large angular momentum $J$ are $0.5$ and approximately $0.25$, respectively~\cite{Herzberg1989}. This gives $\|d\|\approx 4.2\,e a_{0}$, where $e$ is the electron charge and $a_0$ is the Bohr radius. 

% The exact formulas for the H\"{o}nl--London factor for the case of a perpendicular molecular transition  $\Delta \Lambda = \Lambda' - \Lambda'' = +1$ are~\cite{Hansson2005}
% \begin{equation}
% \begin{aligned}
% f_{HL}^{R} & = \frac{(J'' +2 + \Lambda'') (J'' +1 +\Lambda'')}{4(J''+1)}, \\
% f_{HL}^{Q} & = \frac{(J'' - \Lambda'') (J'' +1 +\Lambda'')(2J''+1)}{4 J'' (J''+1)}, \\
% f_{HL}^{P} & = \frac{(J'' - 1 - \Lambda'') (J'' - \Lambda'')}{4 J''},
% \end{aligned}
% \end{equation}
% where $\Lambda''$ is the quantum number that describes the projection of the electronic angular momentum of the molecule onto its internuclear axis in the ground state of the molecule. There are different normalization conventions for the H\"{o}nl--London factors (see Ref.~\cite{Hansson2005}). To use the H\"{o}nl--London factors as a part of a branching ratio, we need to normalize them as
% \begin{equation}
% \tilde{f}_{HL}^{i} = \frac{f_{HL}^{i}}{f_{HL}^{R}+f_{HL}^{Q}+f_{HL}^{P}}.
% \end{equation}
% This is the reduced Hönl--London factor used in~\eqref{eq1}.

% Taking all of this together, for an intensity $I = 100$ mW/cm$^2$ we have Rabi frequency $\Omega_R\approx 127$ MHz, where $\Omega_{R} = \|d\|E/\hbar$, and we have used
% \begin{equation}
% \frac{e a_{0}}{h} = 1.28\,\frac{\rm{MHz}}{\rm{V/cm}}
% \end{equation}
% and
% \begin{equation}
% I\;[{\rm mW/cm^{2}}] = 1.33\,(E\;[{\rm V/cm}])^{2}.
% \end{equation}

In the following, we present rough estimates of the required laser parameters that may be used for a ``first-path'' evaluation of the feasibility of experimental realizations of the various schemes presented above. A more detailed analysis will be required for specific experimental designs. We assume that we have a molecular beam with a velocity $3 \times 10^4$ cm/s. This molecular beam passes two overlapping laser beams in a conventional STIRAP configuration~\cite{Bergmann1998}. If the overlapping beam region has a length of $0.1$ cm, the transit time for a molecule to cross the laser beam is $\tau_t=\SI{3}{\us}$; the size of the overlap region is one of the parameters to be optimally chosen for an experiment.

\emph{Estimate for folding method with $J=100\to J'=100$ transition.} All of the $\log_22J$ iterations of the optical pumping procedure must be completed in $\tau_t$, which requires about $2J=200$ STIRAP transfers in total. The time taken by each transfer operation must be longer than the time $\tau_p$ appearing in the adiabatic condition, which corresponds to just the overlap time between the pump and Stokes pulses. Assuming that the total pulse time is five times longer than the overlap time, with $\tau_p=\tau/100=\SI{0.1}{\ns}$ and using $\tau_p\Omega_R>10$, we get a requirement of $\SI{20}{\kW/\cm^2}$, or $\SI{200}{\W}$ in a $\SI{1}{\mm^2}$ beam. 

\emph{Estimate for folding method with $J=100\to J'=100$ transition and a $J=101$ shelving state.} Consider the weaker transition to the shelving state. We need to have about $2J$ pulses in a transit time of the molecules across the laser beam. Assuming a safety factor of 2, we have $\tau_p=\SI{7}{\ns}$. Including another factor of 10 to the adiabatic condition for the detuned fields gives a required intensity of $\SI{3}{\kW/\cm^2}$, or $\SI{30}{\W}$ in a $\SI{1}{\mm^2}$ beam.

\emph{Estimate for the two-shelving-state method.} We need $2J$ spontaneous decays. Assuming each takes three times the excited-state lifetime, the transit time needs to be $600\tau_0=\SI{7}{\us}$, plus the time needed for the coherent transfers. 

These parameters appear to be within the limits of experimental feasibility, especially since the requirements for the laser power may be relaxed by the use of multipass techniques.

\section{Conclusions}

We have considered general techniques for increasing the efficiency of optical pumping for high-angular-momentum systems, such as diatomic molecules in states with high rotational excitation, in which the upper-state of the transition used for optical pumping is ``leaky,'' i.e., the branching ratio of the spontaneous decay to the target manifold is less than unity.

The goal of the method is to minimize the number of spontaneous-emission events as much as allowed by the second law of thermodynamics as applied to optical pumping.  This is accomplished by arranging the atomic population distribution, prior to spontaneous emission, using stimulated processes such as AFP and/or STIRAP. One variant of the method ``folds'' the atomic populations in half in each iteration; it is not the most efficient method possible, but it requires only approximately $\log_2 2J$ sequential spontaneous decays. Another variant that uses two shelving states has the highest allowed efficiency, with only about one spontaneous decay per atom; this variant requires about $2J$ sequential decays.

We have considered the application of the method to the optical pumping of the K$_2$ molecule. Other molecules may be of interest for application of the method, such as OH (hydroxyl radical), which has recently been actively researched in the context of quantum optics experiments~\cite{Stuhl2012, Stuhl2014}.

The methods described can also be generalized beyond just populating the stretched state. Indeed, once the population has been combined in one state, it can be moved, with an appropriate sequence of coherent population transfers, to any other state, or a coherent superposition thereof. 

\begin{acknowledgments}

The authors gratefully acknowledge stimulating discussions with Hartmut Häffner, Derek F. Jackson Kimball, Thad Walker, Brian Saam, Victor Flambaum, Arne Wickenbrock, Nathan Leefer, Dionysis Antypas, Michael Romalis, Harold Metcalf, and Klaas Bergmann. M.A. acknowledges support from the Taiwanese, Latvian and Lithuanian Research Councils project ``Quantum and Nonlinear Optics with Rydberg-State Atoms'' 2016--2018, FP-20174-ZF-N-100. S.P. acknowledges support from the Polish Ministry of Science and Higher Education within the Iuventus Plus program, and K.S. from the National Centre of Research and Development (the Leader Program).

\end{acknowledgments}

\bibliography{EfficientOP.bib} 

\end{document}